\documentstyle[aps,prl,epsfig,graphicx,twocolumn]{revtex}

\begin{document}
\twocolumn[
\hsize\textwidth\columnwidth\hsize\csname@twocolumnfalse\endcsname
\title{A note on  cluster methods for strongly correlated electron systems}
\author{Giulio Biroli and Gabriel Kotliar}
\address{\it Center for Materials Theory, Department of Physics and
Astronomy, Rutgers University, Piscataway, NJ 08854 USA}

\maketitle
\begin{abstract}
We develop, clarify and test  various aspects of  
cluster methods dynamical mean field methods using
a soluble toy model as a benchmark. 
We find that the Cellular Dynamical Mean Field
Theory (C-DMFT) converges very rapidly  
and compare its convergence properties
with those of the Dynamical Cluster Approximation (DCA).
We   propose  and test  improved estimators for the lattice self
energy within  C-DMFT.
\end{abstract}
]

The development  of dynamical mean field methods has resulted in
significant advances in our understanding of strongly correlated
electron systems, in particular in   the area of the Mott
transition \cite{review}. This method, captures the local
effects of correlations  such as the Kondo effect and the transfer
of spectral weight between the coherent and the incoherent part
of the spectral function.
It suffers however  from  limitations arising
from its single site mean field character such as  
the lack of k dependence of the self energy. 
In the context of disordered system, the dynamical mean field theory has
a precursor in the famous coherent potential approximation (CPA) for which 
cluster extensions such as the molecular CPA \cite{Ducastelle} have been
formulated. Naturally, generalizations of these and other statistical
mechanical approaches such as the Bethe Peirls approximation
to the area of quantum interacting systems,
have been investigated recently
 \cite{review},
\cite{Schiller}, 
\cite{jarrell}
,\cite{katsnelson}
,\cite{edmft}, 
 \cite{Cdmft}. 
This area of investigation is in its beginning stages, and comparative
studies of the various methods are important to increase our
understanding of  the strengths
and the limitations of these methods, at a level comparable to our
present understanding of the single site dynamical mean field theory.
This note is a contribution in this direction.
We focus, here on the CDMFT  \cite{Cdmft} and the DCA \cite{jarrell} method,  
because
both methods have been proved to be manifestly causal, i.e. the
output of an approximated solution of the  cluster equations is causal,
as long as a causal method is used for the solution of the impurity model.
Because no numerical
investigations of the C-DMFT  have  been carried out, 
we test the performance of this method in a simple soluble model
that was introduced by  I. Affleck   and B. Marston \cite{Affleck}.
It has a k dependent, albeit static, self energy, and therefore is
a simple playground to explore the cluster method. 

In the first part of the paper   we describe the   CDMFT \cite{Cdmft},
and introduce a real space formulation of the DCA.
to facilitate the comparison with CDMFT.
In the second part we compare the
predictions of DCA and CDMFT  for the short distance behavior 
of correlation functions, in different cluster sizes
against   the exact solution.
In  the  CDMFT approach
the lattice self energy is a derived quantity which
needs to be 
estimated from the cluster self energy, an auxiliary
quantity  which  enters  
the 
dynamical
mean field equations.
In the second part of the paper we provide 
improved estimators for the lattice self energy and discuss
how they improve the convergence to the exact answer as a function
of the cluster size.

{ \it Real space formulation  the cluster schemes:}
A fairly general   model of strongly correlated
electrons contains hopping and interaction
terms. It is  defined 
on a lattice of $L^{d}$ sites in d dimensions, and we divide the 
lattice in $(L/L_{c})^{d}$ cubic clusters of $L_{c}^{d}$ sites (more general
forms can also be considered). We denote with 
${\mathbf{e}}_{i}$ the internal cluster position and with ${\mathbf{R}}_{n}$ 
the cluster position in the lattice (therefore the position
of the i-th site of the n-th cluster is ${\mathbf{R}}_{n}+{\mathbf{e}}_{i}$).
The lattice Hamiltonian  is expressed in terms of 
fermionic operators $f^{\dagger }_{{\mathbf{R}}_{n},\alpha }$ and $f_{{\mathbf{R}}_{m},\beta }$
and can be written  as:

\begin{eqnarray}\label{hamiltonian}
H&=&\sum_{n,\alpha ,m,\beta }t_{\alpha ,\beta }({\mathbf{R}}_{n}-{\mathbf{R}}_{m})f^{\dagger }_{{\mathbf{R}}_{n},\alpha }f_{{\mathbf{R}}_{m},\beta }\\
&+&\sum_{n,\alpha ,m,\beta ,n',\alpha ',m',\beta '}
U_{\alpha ,\beta ,\alpha ',\beta '}(\{ {\mathbf{R}}\})f^{\dagger }_{{\mathbf{R}}_{n},\alpha }f_{{\mathbf{R}}_{m},\beta }f^{\dagger }_{{\mathbf{R}}_{n'},\alpha '}f_{{\mathbf{R}}_{m'},\beta '}\nonumber
\end{eqnarray}

$\alpha =i,\sigma $ and $\sigma $ is an internal degree
of freedom (i.e. a  spin,   spin-orbital or band index).
Most cluster schemes, neglect the  interaction terms between different clusters.
The effects of those interactions, can be treated using the extended
dynamical mean field approach \cite{edmft} but we will not discuss these
improvements in this paper.
All the different cluster schemes can be formulated as a self consistent 
equation for the cluster self-energy which consist of the  
following loop:(i) start 
with a guess of the cluster self-energy $ (\Sigma _{c})_{\alpha ,\beta }$,
(ii) from the cluster self energy 
compute the Weiss function or 
host cluster propagator $ (G_{0})_{\alpha ,\beta }$, which enters in the effective action for the 
cluster degrees of freedom, (iii) use  the effective action compute the 
cluster Green function $(G_{c})_{\alpha ,\beta }$, (iiii)
compute the new cluster self-energy, (iiiii) iterate this
loop until the convergence is reached. The DCA and CDMFT schemes differ
in the way step (ii) is carried out.
Within the CDMFT one obtains the Weiss function
from the cluster self energy by the equation:
\begin{equation}\label{cdmfteq1}
\hat G_{0}^{-1}=\left(\left(\frac{L_{c}}{L}\right)^{d}\sum_{{\mathbf{K}}}\frac{1}{(i\omega _{n}+\mu)\hat {\cal I} -\hat t ({\mathbf{K}})-\hat \Sigma _{c}}\right)^{-1}+\hat \Sigma_{c} 
\end{equation}
where $t({\mathbf{K}})_{\alpha ,\beta }$ is the Fourier transform of the 
hopping matrix in eq. (\ref{hamiltonian}) 
with respect to ${\mathbf{R}}_{n}-{\mathbf{R}}_{m}$ ($\mathbf{k}$ is a wave-vector in the Brillouin
zone reduced by $L_{c}$ in each direction), $\omega_{n} $ is the Matsubara frequency and $\mu $ is the chemical potential. 
Once the Weiss function has been computed one can obtain $\hat G_{c}$ by functional
integration of the single site action. Step (iiii) is carried 
out using the definition of the cluster self-energy: $\hat \Sigma _{c}=\hat G_{0}^{-1}-\hat G_{c}^{-1}$. 

To facilitate the comparison with the CDMFT,
in the following we shall derive the DCA scheme 
using the  real space formulation  of the cluster. 
To lighten the notation we will  assume that  
the variable $\sigma $ is conserved to make   all the cluster matrices 
diagonal in $\sigma $ and subsequently  we will drop this index.
We take periodic boundary condition on the cluster and we define the matrix $U_{i,j}({\mathbf{K}})
=\exp (-i {\mathbf{K}}\cdot {\mathbf{e}}_{i})\delta _{{\mathbf{e}}_{i},{\mathbf{e}}_{j}}$.
It is crucial for the following to note that the matrix $\hat t({\mathbf{K}})$ has the following representation: $ t({\mathbf{K}})_{i,j}=\frac{1}{L_{c}^{d}}\sum_{{\mathbf{k}_{c}}}e^{i({\mathbf{K}}+{\mathbf{k}}_{c})\cdot ({\mathbf{e}}_{i}-{\mathbf{e}}_{j})}t({\mathbf{K}}+{\mathbf{k}}_{c})$,
where ${\mathbf{k}}_{c}$ are the cluster momenta. Therefore the matrix $\hat U ({\mathbf{K}})\hat t ({\mathbf{K}}) \hat U ^{\dagger }({\mathbf{K}})$ is diagonal with respect to cluster momenta: $(\hat U ({\mathbf{K}})\hat t ({\mathbf{K}})\hat U ^{\dagger }({\mathbf{K}}))_{i,j}=\frac{1}{L_{c}^{d}}\sum_{{\mathbf{k}_{c}}}e^{i{\mathbf{k}}_{c}\cdot ({\mathbf{e}}_{i}-{\mathbf{e}}_{j})}t({\mathbf{K}}+{\mathbf{k}}_{c})$.
Using this property one can write the (ii) DCA eq. in real space as:
\begin{equation}\label{dcaeq1}
\hat G_{0}^{-1}=\left(\left(\frac{L_{c}}{L}\right)^{d}\sum_{{\mathbf{K}}}\frac{1}{(i\omega _{n}+\mu)\hat {\cal I} -\hat U\hat t \hat U ^{\dagger }({\mathbf{K}})-\hat \Sigma _{c}}\right)^{-1}+\hat \Sigma _{c}
\end{equation}
Since the matrices $\hat U\hat t \hat U ^{\dagger }(K) $
and $ \hat \Sigma _{c}$ are
diagonal with respect to ${\mathbf{k}}_{c}$, this equation coincides with
the  DCA equations  of Jarell et. al. \cite{jarrell} after  a Fourier transformation with cluster
momenta.
Once $\hat G_{0}$ is known, $\hat G_{c}$ is computed
by functional integration of the cluster effective action and the 
new cluster self-energy is obtained by $\Sigma _{c}({\mathbf{k}}_{c})=G_{0}^{-1}({\mathbf{k}}_{c})-G_{c}^{-1}({\mathbf{k}}_{c})$.\\
Eq.  \ref{dcaeq1}  allows a direct formulation of DCA in real
space and a detailed comparison with CDMFT. We  also note this
real space formulation can be used to defined many causal cluster
schemes, by introducing  a different  matrix $ U(k)$ in the previous equation. 

{\it A simplified one dimensional large-N model:
comparison between the exact solution and the predictions of the cluster schemes.}
In the following we focus on a simple one dimensional model,
originally introduced and studied by  I. Affleck and B. Marston \cite{Affleck} in two dimensions.
We compare the DCA and CDMFT schemes to its exact solution. 
This model is  a generalization
of the Hubbard-Heisenberg model where the SU(2) spins are replaced by
a  SU(N) spins, the on site repulsion is scaled as $1/N$ and the large N limit
is taken. Its Hamiltonian reads:
\begin{eqnarray}\label{hamiltoniansimple}
H&=&-t\sum_{i,\sigma }(f_{i,\sigma }^{\dagger}f_{i+1,\sigma }
+f_{i+1,\sigma }^{\dagger}f_{i,\sigma })\\
&+&\frac{J}{2N}\sum_{i,\sigma ,\sigma '}(f_{i,\sigma }^{\dagger}f_{i,\sigma' }f_{i+1,\sigma '}^{\dagger}f_{i+1,\sigma}+f_{i+1,\sigma }^{\dagger}f_{i+1,\sigma' }f_{i,\sigma '}^{\dagger}f_{i,\sigma})\nonumber
\end{eqnarray}
\begin{figure}[bt]
\centerline{    \epsfysize=7cm
       \epsffile{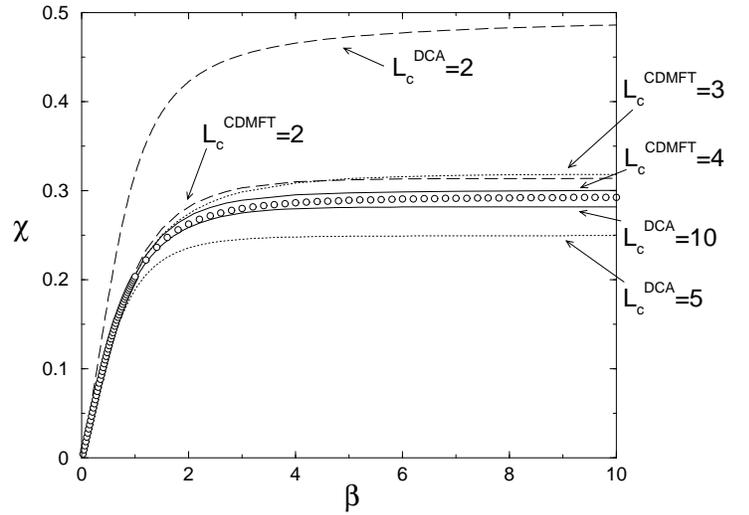}}
\caption{$\chi$ as a function of $\beta $ for $\mu =1$, and $t=1$. 
The points are the exact solution. 
The lines are $\chi _{cl}^{DCA}$ for $L_{c}=2,5,10$ and $\chi _{cl}^{CDMFT}$ for $L_{c}=2,3,4$.
\label{conv.fig}}
\end{figure}

where $i=1,\dots ,L$ and $\sigma =1, \dots , N$ and we take the large $L$
and $N$ limits. In the following we will use $J$ as the unite of temperature 
and therefore we put $J=1$ and we rescale the hopping term $t\rightarrow t/J$.
The thermodynamics of this model can be solved exactly 
since in the large $N$ limit the quantity $ \chi = \frac{1}{N}\sum_{\sigma }f_{i,\sigma }^{\dagger}(t)f_{i+1,\sigma}(t)$ does not fluctuate. Indeed
(\ref{hamiltoniansimple}) reduces to a free-fermions
Hamiltonian with a ``renormalized'' hopping term $t\rightarrow t+\chi $ 
and a self-consistent condition on $\chi $:
\begin{equation}\label{selfsimple}
\chi =\frac{1}{L}\sum_{k}f(\beta E_{k})\cos k, \qquad E_{k}=-2(t+\chi )\cos k +\mu 
\end{equation}
where $\mu$ is the chemical potential, $f(\beta E_{k})$ is the Fermi function
and $\beta $ is the inverse temperature.\\
We now apply the DCA approximation to the Hamiltonian 
(\ref{hamiltoniansimple}). 
As previously, the computation is simplified by the fact that 
the quantity $ \chi _{cl}^{DCA}= \frac{1}{N}\sum_{\sigma }f_{i,\sigma }^{\dagger}(t)f_{i+1,\sigma}(t)$, where $i$ and $i+1$ belong to the same cluster, 
does not fluctuate in the large-N limit. As a consequence 
the functional integral on the cluster degrees of freedom 
reduce to a simple Gaussian integral. Thus, imposing periodic boundary 
condition on the cluster, the eq. ${\cal G}^{-1}(k_{c})=G_{c}^{-1}(k_{c})
+\Sigma _{c}(k_{c})$ implies: $\Sigma _{c}(k_{c})=2\chi _{cl}^{DCA}\cos k_{c}$.
Using  the second DCA equation which  expresses the cluster Green-function 
as a function of $\Sigma _{c}(k_{c})$,
we  obtain the self-consistent DCA relation for  $\chi_{cl}^{DCA} $:
\begin{eqnarray}\label{eqchi}
\chi _{cl}^{DCA}&=& \frac{1}{L}\sum_{K,k_{c}}f(\beta E_{K,k_{c}})\cos k_{c}, \\
&&\qquad E_{K,k_{c}}=-2t\cos (k_{c}+K) -2\chi _{cl}^{DCA}\cos k_{c}+\mu \nonumber
\end{eqnarray}
Note that in the infinite cluster limit one recovers the exact equation (\ref{selfsimple}).
\begin{figure}[bt]
\centerline{    \epsfysize=7cm
       \epsffile{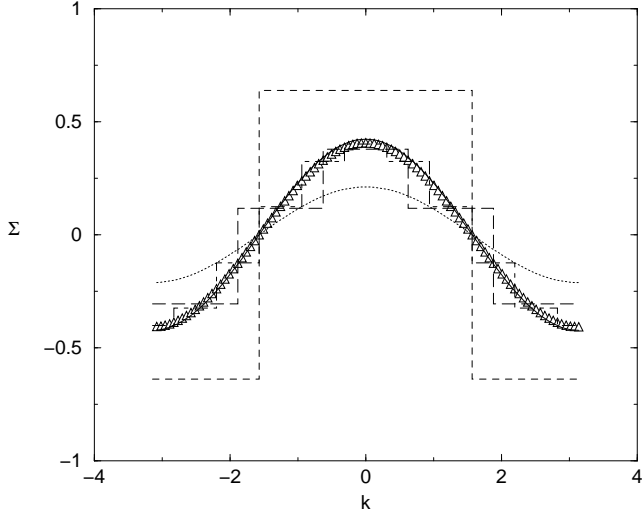}}
\caption{Lattice self energy predicted by the different methods compared
with the exact solution (triangles) for $t=\beta =\mu =1$. The dotted line is the result of the CDMFT estimator  in [6], 
whereas the continuous line is the
result of the first improvement discussed in this paper. The dashed, long dashed and dot dashed lines are respectively the DCA results for $L_{c}=2,5,10$. 
\label{self.fig}}
\end{figure}

Now we focus on the CDMFT approximate solution. As in the DCA case, since 
the quantity $ (\chi_{cl}^{CDMFT}) _{i}= \frac{1}{N}\sum_{\sigma }f_{i,\sigma }^{\dagger}(t)f_{i+1,\sigma}(t)$ does not fluctuate, one obtains:  $(\Sigma _{c})_{i,j}=(\chi_{cl}^{CDMFT})_{i}\delta _{i,j-1}+(\chi_{cl}^{CDMFT}) _{j}\delta _{i,j+1}$.
This is the generalization of the corresponding DCA expression to a case
without periodic boundary condition. Note that now the quantity 
$(\chi_{cl}^{CDMFT})_{i}$ may depend on the cluster index.
Denoting the eigenvectors and the eigenvalues of the matrix 
$\hat t + \hat \Sigma  $ respectively $\psi _{i}^{\nu }(K)$ and $\lambda ^{\nu }(K)$ ($\nu =1,\dots ,L_{c}$), the (ii) CDMFT equation, which expresses 
the cluster Green-function in terms of the cluster self-energy,
reads:
\begin{equation}\label{clgf}
(G_{c})_{i,j}=\frac{L_{c}}{L}\sum_{K,\nu }\psi _{i}^{\nu }(K)(\psi _{j}^{\nu }(K))^{*}\frac{1}{i\omega _{n}+\mu-\lambda ^{\nu }(K)}
\end{equation}
Using this expression we finally get the self-consistent CDMFT 
equation on $(\chi_{cl}^{CDMFT}) _{i}$:
\begin{eqnarray}\label{selfDCMFT}
(\chi_{cl}^{CDMFT}) _{i}&=&\frac{L_{c}}{L}\sum_{K,\nu }\psi _{i}^{\nu }(K)(\psi _{i+1}^{\nu }(K))^{*}f(\beta E_{K,\nu }),\qquad\\
&& E_{K,\nu }=\mu-\lambda ^{\nu }(K)\nonumber
\end{eqnarray}
Notice  that  eq. \ref{selfDCMFT} corresponds to the exact solution of a
model defined by a Hamiltonian similar to (\ref{hamiltoniansimple}) 
in which $J_{i,i+1}$ equals $1$ if $i$ and $i+1$ belong to the same cluster
and zero otherwise. This implies in particular that in the infinite cluster 
limit the CDMFT approximation gives back the exact solution.
We have numerically solved the self consistent equations
 (\ref{selfsimple},\ref{eqchi},\ref{selfDCMFT})
 to compare the DCA and CDMFT predictions for different cluster
sizes  to the exact solution. 
\begin{figure}[bt]
\centerline{    \epsfysize=7cm
       \epsffile{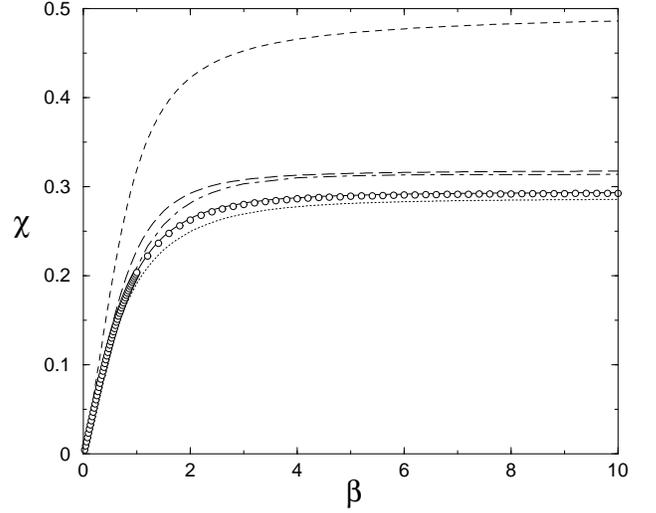}}
\caption{$\chi$ as a function of $\beta $ for $\mu =1$, and $t=1.$. 
The points are the exact solution. 
The lines are, from top to bottom: $\chi _{cl}^{DCA}$ (dashed line), $\chi_{la}^{DCA}$ (long-dashed line), $\chi _{cl}^{CDMFT}$ (dot-dashed line),
$\chi_{la}^{CDMFT2}$ (continuous line), $\chi_{la}^{CDMFT1}$ (dotted line). $\chi_{la}^{CDMFT1}$ is the result obtained using the CDMFT estimator for the lattice self-energy in [6] whereas $\chi_{la}^{CDMFT2}$ corresponds to the first improvement discussed in this paper.
\label{t1.fig}}
\end{figure}

In Fig. (\ref{conv.fig}) we plot the result of this analysis 
for $t=1,\mu =1$ as a function of $\beta $ and 
for different cluster sizes. The two methods converge (the convergence
is   not uniform  in $\beta $)
toward the exact solution for high enough $L_{c}$ but
C-DMFT converges better that DCA. Indeed 
the C-DMFT result are already surprisingly good for $L_{c}=2$.
However, as we shall discuss below,
there are two different ways to compute $\chi$ within the cluster methods. 
The one used  here is based on a real space cluster intuition. The second
one, relies on a momentum space intuition and computes the correlation functions
from the k dependent lattice Greens function, this procedure was advocated
in ref \cite{jarrell} and .
we will show that 
it is also gives  accurate    results for small clusters.

{\it The lattice self-energy.} We now address
the computation of 
the
lattice self  energy.
In   DCA   a discretized form of  the  lattice self energy  in momentum
space, enters directly in the evaluation  of $G_0$. On the other hand,
CDMFT focuses   on estimating the  cluster greens function,  and the
lattice self energy does not participate in the mean field equations,
and has to be estimated later from the cluster self energy.
For the simplified large-N one dimensional model studied in this paper the
DCA prediction for the lattice self-energy reads:
\begin{equation}\label{selfDCA}
\Sigma _{latt}^{DCA}(k)=\Sigma (k_{c})=2\chi _{cl}^{DCA}\cos k_{c} 
\end{equation}
where $k$ belongs to $[-\frac{\pi }{L_{c}}+k_{c},k_{c}+\frac{\pi }{L_{c}}]$. Whereas for C-DMFT
an   estimator for the  lattice self energy is
constructed  using  the matrix $S_{R_{n},\alpha ; i}=\delta _{R_{n}+\alpha ,i}$ where $\alpha $ 
is the internal cluster index and $i$ denotes a lattice site.
The simplest form \cite{Cdmft} is:
$\Sigma _{latt}^{CDMFT}(k)=\sum_{\alpha ,\beta }\tilde{S}^{\dagger }_{\alpha }
(k)\Sigma _{\alpha ,\beta }\tilde{S}_{\beta }
(k)$,
where $\tilde{S}_{\alpha }(k)$ is the Fourier transform of the matrix $S$
with respect to the original lattice index $i$.
$\tilde{S}_{\alpha }(k)$ can be easily written in 
terms of the matrix $\hat  U(k)$ defined before: $\tilde{S}(k)_{\alpha }=U_{\alpha ,\alpha }(k)/\sqrt{L_{c}}$.
Therefore the relationship between the lattice and the cluster self energy
reads: $\Sigma _{latt}^{CDMFT}(k)=\sum_{\alpha ,\beta }[U^{\dagger }(k)\Sigma_{c}U(k)]_{\alpha ,\beta }/L_{c}$.
For example, in the case of the two site cluster we find: $\Sigma _{latt}^{CDMFT}(k)=\chi_{cl}^{CDMFT} cos(k)$, whereas the exact solution gives $\Sigma _{latt}(k)=2 \chi_{ex} cos(k)$.  As a consequence, even if the value of $\chi$ is well predicted by the CDMFT there
is a factor 2 between the two self-energies. The reason of this discrepancy
may be understood writing the simple estimator of the lattice self energy  \cite{Cdmft}
in real space $(\Sigma _{latt})_{i-j}=\sum_{\alpha ,\beta  \,:\,
\alpha -\beta =i-j}(\Sigma_{c})_{\alpha ,\beta }/L_{c}$.
This means that the lattice self energy for a certain value of 
$i-j$ is obtained
averaging over all the cluster self energy elements corresponding to
$\alpha -\beta =i-j$. In the limit of an infinite cluster  translation
invariance implies that the cluster self energy coincides
with the lattice self energy in the bulk.
Therefore the factor $1/L_{c}$ cancels and
we get the exact solution.
However for a finite lattice there are only $L_{c}-1$ factors for $i-j=1$,
 $L_{c}-2$ factors for $i-j$=2 , \dots $L_{c}-k$ factors for $i-j=k$.
Therefore it is highly desirable  to have improved estimators for
smaller size clusters in which the  formula in which the average over 
all the factors having $\alpha -\beta =k$ is weighted by their number $1/(L_{c}-k)$.
One could also think to put an extra weight to extract the lattice self-energy
only from the sites in the bulk, for which the CDMFT result should be better.
We propose  new general class of estimators for the lattice self energy
in terms of the cluster self energy, that inherit  its  causality
property. 
\begin{equation}\label{estim}
(\Sigma _{latt})_{i-j}=\sum_{\alpha ,\beta \,:\,
\alpha -\beta =i-j}w_{\alpha ,\beta }{(\Sigma_{c})_{\alpha ,\beta }}
\end{equation}
where the matrix $w_{\alpha ,\beta }$ is positive definite and
$\sum_{\alpha ,\beta \,:\, \alpha -\beta =i-j}w_{\alpha ,\beta }\rightarrow 1$ 
for $L_{c}\rightarrow \infty $ (this guarantees a good behavior in the infinite
cluster limit). Using that the trace of the product of two positive definite
matrices is positive, one can easily prove that if 
the cluster self-energy is causal this formula produce a lattice 
self-energy which is also causal. Note that (\ref{estim}) does not change the behavior 
for an infinite cluster, but can really improve the results for finite cluster
sizes. For example, in Fig. \ref{self.fig} we compare the exact lattice 
self-energy to the DCA and the CDMFT predictions 
(using the initial estimator  proposed in \cite{Cdmft}
and the simple improvement $w_{\alpha ,\beta }=1/(L_{c}-1)$ which weights in the right
way at least the terms with $\alpha -\beta =1$) for $\beta =\mu =t=1$.
We remark 
that there is an excellent 
agreement between CDMFT and the exact solution after that our  simple
improvement
has been taken into account already for $L_{c}=2$. Whereas the DCA prediction 
becomes good for $L_{c}\geq 5$.

{\it Relation between lattice and cluster observables.}
Once  the lattice self-energy has been obtained within a cluster method,
the lattice Green function can be  straightforwardly computed.  This
offers a different way of estimating 
the first neighbor correlation function $\chi$, using the lattice
Greens function. 
This quantity can be computed inside the cluster ($\chi_{cl}$) 
or using the lattice Green function, obtained by the lattice self-energy,
($\chi_{la}$) and the two
results do not coincide in general. In the case of C-DMFT one can understand 
what are the approximations responsible for  this difference and why they
are small. The C-DMFT approach is based 
on the cavity procedure \cite{Cdmft} which, if it was carried out exactly, it would give back the
same answer for the lattice and cluster observables. However, in the approximated cavity procedure
adopted in the C-DMFT, one assumes that the contribution to the effective action coming from 
tracing out all the degrees of freedom outside the cluster is purely gaussian. This is clearly not the
case in general and it is the main reason for the non-coincidence of lattice and cluster observables.

In Fig. (\ref{t1.fig}) we compare the DCA and CDMFT predictions for the lattice
and cluster values of $\chi$ to the exact solution for a two site cluster, for 
$t=1,\mu =1$ as a function of $\beta $.   
These curves display the typical behavior found also for other values of the control
parameters: $\chi_{la}^{DCA}$ is quite better than its cluster counterpart,
whereas the C-DMFT prediction is quite stable. This is probably 
the result of having an approximate 
cavity construction for the C-DMFT \cite{Cdmft}. Moreover 
we remark that $\chi_{la}^{CMDT2}$, obtained using the
first improvement for the self-energy discussed above, almost coincides with the exact solution.
Comparing the C-DMFT and the DCA lattice values of $\chi$ we note that C-DMFT gives usually a little bit better answer in terms of accuracy and convergence with respect to cluster size. 
This is perhaps due to the smoothness of the self-energy in the C-DMFT case. 

It would be nice to eliminate the cluster self energy altogether
from the C-DMFT approach or to use a self energy without discontinuities
in DCA, 
in the  spirit of the work  of Katsnelson and Lichtenstein  
\cite{katsnelson}. However, we were unable to prove manifest causality of
this approach. 

In summary, in 
this short note we compared the performance of the DCA method with
that of the cellular DMFT, in a  very simple  toy  model.
We have also 
proposed  new estimators for the lattice self energy
within CDMFT, which is     
more efficient. 
Our study  shows that  a direct application of CDMFT, 
i.e.  without exploiting the flexibility inherent in the
choice of basis in its most general formulation  is very efficient
in converging to the correct solution already for a two site cluster. 
Comparing the  DCA and  the C-DMFT predictions are a little bit better
in terms of accuracy and convergence with respect to cluster size.
DCA estimates of physical quantities, are most accurately carried
out using the lattice Greens function, and not from the real space cluster
correlation functions. This  is stressed in ref \cite{jarrell},  who view
DCA as a momentum space method. C-DMFT  is not so sensitive to the choice
of cluster or lattice estimators, because of the underlying a cavity
construction of the derivation \cite{Cdmft}.    
These results are very encouraging, and warrant further applications
of these methods to more realistic  and difficult problems. 
Since the  most glaring deficiency of the CDMFT method is that
it does not attempt to take into account in a direct fashion the
translation invariance of the  problem, we concentrated on
the phase of this model (finite temperatures) 
which is translationally invariant. We 
expect that CDMFT will perform even better for the ground state properties
since in this case translation invariance is broken by dimerization.

Acknowledgments:
This work was supported by the NSF under grant DMR-0096462
and by  the Rutgers Center for Materials Theory.
GK acknowledges useful discussions with A. Georges,   M. Jarrell, H. Krishnamurthy
and Sasha Lichtenstein.

\end{document}